\documentclass[12pt]{article}

\usepackage{graphicx}

\newcommand{\apjl}{Astrophys. J. L.}
\newcommand{\apj}{Astrophys. J.}
\newcommand{\apjs}{Astron. J. Supplement Series}

\newcommand{\aap}{Astron. Astrophys.}

\newcommand{\mnras}{Mon. Not. R. Astron. Soc.}

\newcommand{\pasp}{Pub. Astron. Soc. Pacific}
\newcommand{\aspcs}{Astron. Soc. Pacific Conference Series}
\newcommand{\spieconf}{Society of Photo-Optical Instrumentation Engineers (SPIE) Conference Series}
\newcommand{\science}{Science}

\makeatletter
\def\@cite#1#2{$^{\mbox{\scriptsize #1\if@tempswa , #2\fi}}$}
\makeatother

\title{In the Shadow of the Transiting Disk: Imaging epsilon Aurigae in Eclipse}

\author{
Brian Kloppenborg$^1$, Robert Stencel$^1$, John Monnier$^2$, Gail Schaefer$^3$, \\
Ming Zhao$^4$, Fabien Baron$^2$, Hal McAlister$^3$, Theo ten Brummelaar$^3$, \\
Xiao Che$^2$, Chris Farrington$^3$, Ettore Pedretti$^5$, PJ Sallave-Goldfinger$^3$, \\
Judit Sturmann$^3$, Laszlo Sturmann$^3$, Nathalie Thureau$^5$, \\
Nils Turner$^3$, Sean M. Carroll$^6$
}

\begin{document}

\maketitle

$^1$ University of Denver.
$^2$ University of Michigan.
$^3$ CHARA/Georgia State University.
$^4$ JPL.
$^5$ University of St. Andrews, Scotland, UK.
$^6$ California Institute of Technology.

\textbf{
Astrophysical objects with unusually high mass to luminosity ratios attract attention. The fifth brightest star in the constellation Auriga, $\epsilon$ Aurigae, is such a case. The companion in this single-line spectroscopic binary has evaded direct detection for over 175 years \cite{2002ASPC..279..121G, 1985eepa.rept.....S}. For the first time, closure-phase interferometric imaging has directly detected the eclipsing body, allowing us to measure the properties of the companion. We used the MIRC four-telescope beam combiner at the CHARA Array to obtain images of epsilon Aurigae during ingress into eclipse during autumn 2009. These images show the intrusion of a dark, elongated structure that resembles the large disk as first discussed by Ludendorff \cite{1912AN...L}, extended by Kopal\cite{1954TO...K}, modeled by Huang\cite{1965ApJ...141..976H}, and inferred with polarimetry obtained by Kemp\cite{1986ApJ...300L..11K} . These observations deﬁnitively exclude alternate models suggested by Str\"{o}mgren\cite{1937ApJ....86..570K} and Hack\cite{1964MSA...H} and provide important constraints on the geometrically thin, optically thick disk size, mass, and scale height. In essence, we have observed the beginning of a 18-month long partial solar eclipse, 2000 light years away.
}

Data were collected using Georgia State University's Center for High Angular Resolution Astronomy (CHARA) interferometer\cite{2005ApJ...628..453T} using the Michigan Infra-Red Combiner (MIRC)\cite{2006SPIE.6268E..55M}.  The CHARA Array is located atop Mount Wilson, CA and consists of  six 1-$m$ telescopes capable of 15 baselines ranging from 34 $m$ to 331 $m$.  The longest baseline provides resolutions up to 0.5 $mas$ (milliarcseconds, 28 nano-degrees) at H-band ($\lambda$ 1.50-1.74 $\mu m$).  

All data presented in this Letter were collected during the start of the 2009-2011 eclipse on Nov. 2-4 and Dec. 2-4 (see Supplementary Table 1 for details).  Each four telescope configuration provides six visibilities, four closure phases, and four triple amplitudes simultaneously in each of MIRC's eight narrow spectral channels across the H-band.  There were four pre-eclipse observations during 2008 Nov. - Dec. that verified the F-star is not highly asymmetric and aided in planning the 2009 observations.  These data are consistent with the results obtained at PTI\cite{2008ApJ...689L.137S} (a K-band, $\lambda$ 2.00 - 2.38 $\mu m$, uniform disk diameter of 2.27 $\pm$ 0.11 $mas$) and therefore are not rediscussed in this Letter.

The data from MIRC were reduced and calibrated against seven calibration stars (see Supplementary Table 2) using the standard reduction pipeline\cite{2007Sci...317..342M}, producing nightly OIFITS\cite{2005PASP..117.1255P} data files.  Files from each of the three consecutive nights of observation in 2009 were merged to produce a single OIFITS file for each sequence of observations.  The resulting UV plane, power spectrum, and closure phase coverage (see Figure 1 for interpolated UV plots) is arguably the most complete in optical interferometry to-date.

The 2009 data provided ample UV coverage for interferometric imaging.  Image reconstruction for the figures presented in this Letter were performed using the BiSpectrum Maximum Entropy Method (BSMEM)\cite{2008SPIE.7013E.121B, 2006SPIE.6268E..59L} and Markov Chain Imager (MACIM)\cite{2006SPIE.6268E..58I} software packages.  While the theory behind image reconstruction is common to these packages, based on the minimization of the $\chi^2$ plus a regularization function, they use significantly different approaches to achieve it: a local gradient-based approach and global stochastic minimization by simulated annealing, respectively.  Despite the differences in implementation\cite{2007Sci...317..342M, 2008ApJ...684L..95Z}, the images produced by these packages are in remarkable agreement.  This is proof of the soundness and reliability of the reconstructed images and because of this fact, we present only the MACIM images in this Letter.

Figure 2 shows the 2009Nov and 2009Dec observations in which we see a single object with a circular outline that is notably darker in the South-East quadrant.  In the December image, the overall size of the dark region has grown, but the size of the circular object has remained nearly the same.  The Northern hemisphere of the circular object shows variations at the 15\% level that we believe approximate our photometric errors.  The obscuration in these images was not seen or implied by our previous data sets and we interpret this object to be the theorized disk in the system.  The compactness of the disk across the two epochs provides the first direct evidence that the disk is geometrically thin, but optically thick.

We attempted to model the obscuration using parabolas, ellipses, and rectangles with and without smoothed edges using the reconstructed images as a guide.  A satisfactory fit to the visibilities and closure phases was obtained by using a smoothed-edge obscuring ellipse whose semi-major axis was fixed to 6.10 $mas$ (based on eclipse timing and well-known orbital parameters\cite{2010:Hoard, 1996ApJ...465..371L}), combined with a power-law limb-darkened stellar component.  The nine remaining parameters (stellar diameter, stellar limb-darkening coefficient, ellipse semi-minor axis, ellipse position angle, ellipse smoothing coefficient and ellipse centroid (x,y) for both 2009Nov and 2009Dec) were simultaneously fit using a Levenberg-Marquardt least-squares minimization algorithm.  The combined fit had a reduced $\chi^2$ of 4.69 and predicted a F-star Limb-darkened diameter of 2.41 $\pm$ 0.04 $mas$, or uniform-disk diameter of 2.10 $\pm$ 0.04 $mas$.  The semi-minor axis for the ellipse is 0.61 $\pm$ 0.01 $mas$.  The ellipse has a position angle of 119.80 $\pm$ 0.74 degrees with a full width half max smoothing length of 0.38 $\pm$ 0.05 $mas$.  Between 2009Nov and 2009Dec, the position of the ellipse centroid moved 0.62 $\pm$ 0.14 $mas$ West and 0.34 $\pm$ 0.06 $mas$ North.   

Although the $\chi^2$ is high, the model does an excellent job of reproducing the observed drop in H-band flux (0.40 and 0.53 mag for 2009Nov and 2009Dec respectively), therefore we consider the model a good approximation for the leading edge of the disk.  The ellipse model implies a North-West motion between the two images that lies along a line with a position angle of 296.82 $\pm$ 6.85 degrees.  In the limiting case that the disk is considered infinitely thin, the above parameters imply that the disk has a minimum inclination, $i$, of 84.30 $\pm$ 0.15 degrees.  

Adopting the Hipparcos-estimated distance \cite{1997A&A...323L..49P} of 625 pc (parsec) we may derive estimates of the disk’s physical extent. If the disk were to be viewed edge-on (i.e. $i = 90$) then the maximum thickness of the disk, implied by twice the semi-minor axis of the ellipse, is $0.76 \pm 0.02$ AU (astronomical units). The observed motion of the disk is $0.43 \pm 0.08$ AU , over the observation interval, which implies a relative motion of $25.10 \pm 4.65$ km/s. Using the F-star as the reference point and removing its known semi-amplitude, $15.00 \pm 0.58$ km/s  \cite{1970VA..W}, the velocity of the disk is $10.10 \pm 4.68$ km/s.

These velocities imply the F-star to companion mass ratio is 0.62 $\pm$ 0.12, providing modest evidence that the companion is the more massive component in the system.  Ultraviolet data implies the presence of a hot source inside of the disk which can be fit\cite{2010:Hoard} by a B5V star.  Adopting 5.9 $\pm$ 0.1 $M_{\odot}$ as representative of a B5V star and treating the disk as having negligible mass, we obtain 3.63 $\pm$ 0.68 $M_{\odot}$ for the mass of the F-star.  The mass function\cite{1970VA..W} for the system, 3.12, implies the lower end of the mass range is preferred.

The mass of the disk can be estimated from the volume and a plausible density, if we assume a characteristic near-infrared opacity, $\kappa$, of 10 $cm^2$ $g^{-1}$ and adopt a range of characteristic length scales between the resolution of CHARA to the size of the disk's semi-major axis (i.e. 1.75 $\pm$ 0.87 $AU$), the density of the disk is 3.82 $\pm$ 2.70 $\times$ $10^{-12}$ $kg$ $m^{-3}$.  Modeling the disk as a cylinder with a radius 3.81 $\pm$ 0.01 $AU$ and height 0.76 $\pm$ 0.02 $AU$, the mass of the disk would be 4.45 $\pm$ 3.15 $\times$ $10^{-7}$ $M_{\odot}$, or about 0.15 Earth masses.  For interstellar dust to gas ratios, the total disk mass could rise to 15 Earth masses, 4.45 $\pm$ 3.15 $\times$ $10^{-5}$ $M_{\odot}$, which is far below masses for either stellar component, making the disk mass dynamically negligible.

Direct imaging of the $\epsilon$ Aurigae system has provided validity to the disk model
for the previously unseen companion, first discussed by Kopal\cite{1954TO...K} and modeled by by Huang \cite{1965ApJ...141..976H} and refined with polarimetry by Kemp et. al. \cite{1986ApJ...300L..11K}. With these images and simple model, we can specify the dimensions and masses of the components in the system. In addition, the optically thick but geometrically thin disk is unlike young stellar object disks, but more nearly resembles debris disks. Further imaging will reveal details of the disk structure, density gradients, and scale heights, demonstrating the power of closure-phase imaging when extensive UV coverage is possible.

\newpage

\section*{Acknowledgments}
We are grateful to the firefighters who defended Mount Wilson from the Station fire, and Larry Webster and the staff at Mount Wilson for facilitating our observations. We acknowledge with thanks the variable star observations from the AAVSO International Database contributed by observers worldwide and used in this research.  The CHARA Array, operated by Georgia State University, was built with funding provided by the National Science Foundation, Georgia State University, the W. M. Keck Foundation, and the David and Lucile Packard Foundation. This research is supported by the National Science Foundation as well as by funding from the office of the Dean of the College of Arts and Science at Georgia State University. MIRC was supported by the National Science Foundation.  The University of Denver participants thank Jeff Hopkins for ongoing photometry and are grateful for the bequest of William Hershel Womble in support of astronomy at the University of Denver.

\section*{Author Contributions}
R.S. originally proposed this research task, facilitated observations, and arranged for concurrent observations at other research facilities.  Raw data from MIRC was reduced by J.D.M. using calibrated diameters from X.C. and literature sources.  Image reconstruction and modeling was performed by F.B., J.M., and B.K.  Determination of the F-star's translational velocity was done by G.S. and B.K.. Observations were planned by R.S., G.S., B.K., and M.Z. and was facilitated by J.S., E.P., L.S., N.Thureau, N.Turner, and X.C..  The data necessary for this publication was collected by M.Z., B.K., G.S., C.F., P.S-G., R.S., and F.B..  Discussion of historical models and their implication to observations was conducted by S.M.C., B.K., and R.S..  Administrative oversight and access to CHARA was provided by H.M. and T. tB. All authors discussed the results and commented on the manuscript.

\section*{Author Information}
Reprints and permissions information is available at npg.nature.com/reprintsandpermissions.  Correspondence and requests for materials should be addressed to B.K. (bkloppen@du.edu) or R.S. (rstencel@du.edu)

\newpage

\begin{figure}
\includegraphics[width=\linewidth]{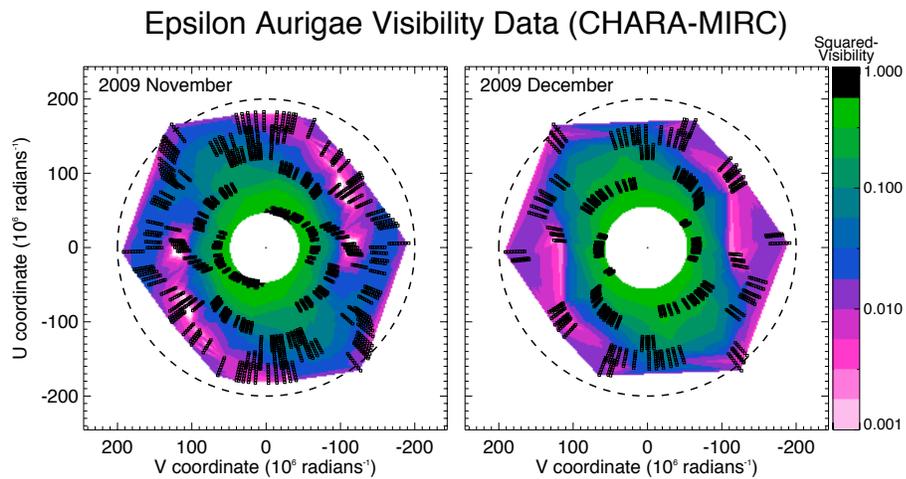}
\caption{Interpolated Visibility Squared (colors) UV plots showing the baseline coverage of the observations (2009 November 2-4, 2009 December 2-4) and a clear asymmetry in the visibility pattern. The dashed circle corresponds to the longest baseline at CHARA (331m) at the middle of the H-band ($\lambda$ 1.65 $\mu$m.) The eﬀective position of the telescopes are shown as squares.}
\end{figure}

\begin{figure}
\includegraphics[width=\linewidth]{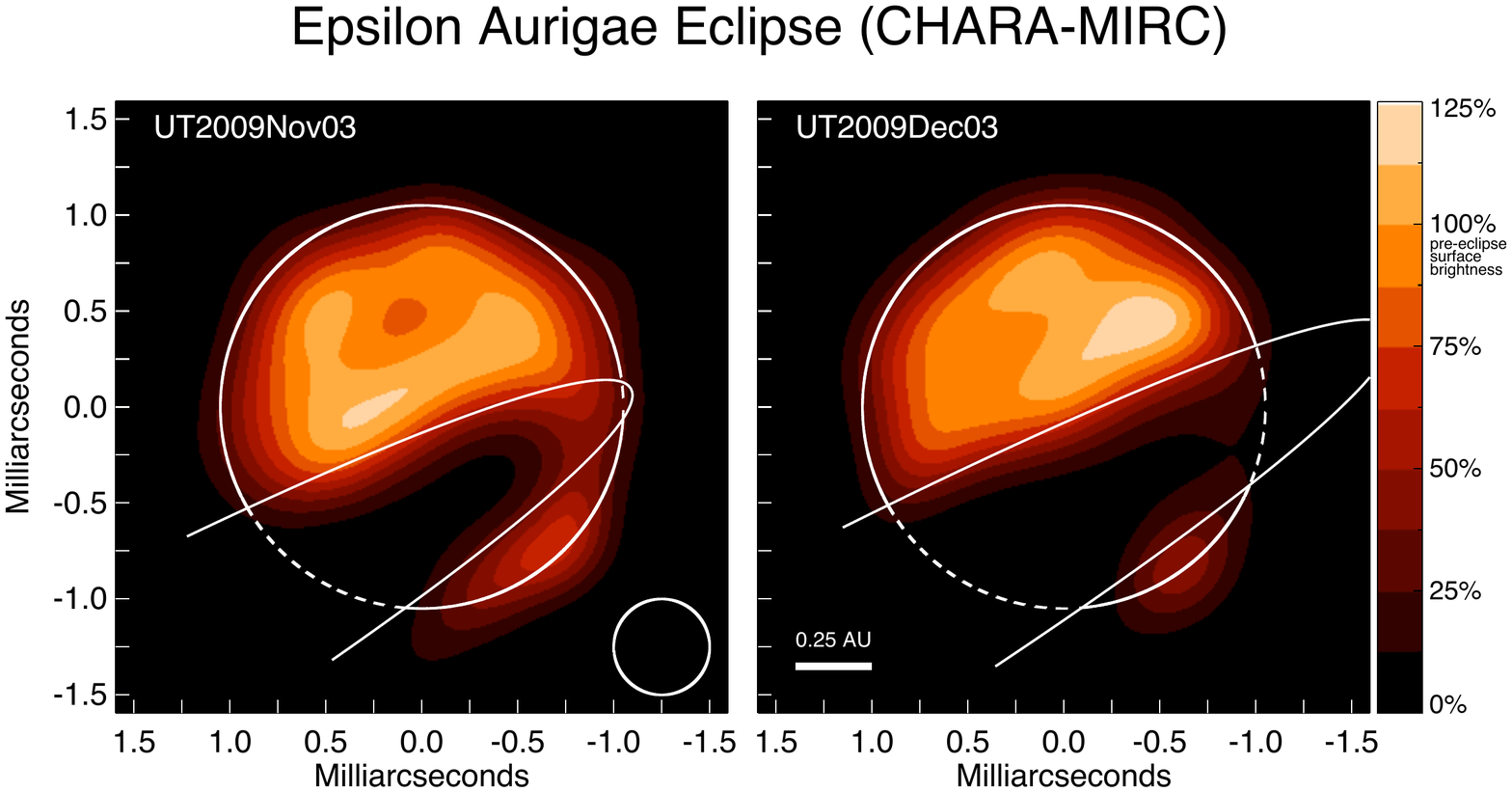}
\caption{The synthesized images from the 2009 Observations. The model discussed in the text is superimposed on the image in white. A circle of diameter of $2.27$ mas is drawn for the F-star and the position of the ellipse for each epoch is shown. CHARA’s H-band resolution ($0.5$ mas) is shown in the bottom right of the left ﬁgure. In order to represent our images in terms of pre-eclipse surfacebrightness, we have assumed eclipse depths of $0.40$ mag and $0.53$ mag for 2009 Nov and Dec respectively based on the ongoing AAVSOmonitoring.}
\end{figure}


\begin{thebibliography}{00}
\bibitem{2002ASPC..279..121G} \textit{Guinan, E.F., Dewarf, L.E.} Toward Solving the Mysteries of the Exotic Eclipsing Binary {$\epsilon$} Aurigae: Two Thousand years of Observations and Future Possibilities \textit{\aspcs} \textbf{279}, 121-142 (2002)

\bibitem{1985eepa.rept.....S} \textit{Stencel, R.E.} The 1982-1984 Eclipse of Epsilon Aurigae \textit{NASA Conference Publications} \textbf{2384}, 1-103 (1985) 

\bibitem{1912AN...L} \textit{Ludendorff, H. } Bearbeitung der Schmidtschen Beobachtungen des Ver\\"{a}nderlichen $\epsilon$ Aurigae. \textit{Astronomische Nachrichten} \textbf{192}, 389–406 (1912) 

\bibitem{1954TO...K} \textit{Kopal, Z} The nature of the eclipses of epsilon Aurigae \textit{The Observatory} \textbf{74}, 14–20 (1954)

\bibitem{1965ApJ...141..976H} \textit{Huang, S.S.} An Interpretation of {$\epsilon$} Aurigae. \textit{\apj} \textbf{141}, 976-984 (1965)

\bibitem{1986ApJ...300L..11K} \textit{Kemp, J.C. et al.} Epsilon Aurigae - Polarization, light curves, and geometry of the 1982-1984 eclipse \textit{\apjl} \textbf{300}, L11-L14 (1986)  
  
\bibitem{1937ApJ....86..570K} \textit{Kuiper, G.P., Struve, O., Str{\"o}mgren, B.} The Interpretation of {$\epsilon$} Aurigae \textit{\apj} \textbf{86}, 570-612 (1937)

\bibitem{1964MSA...H} \textit{Hack, M.} A new explanation of the binary system EPS Aur \textit{Memorie della Societa Astronomica
   Italiana} \textbf{32}, 351–364 (1961)

\bibitem{2005ApJ...628..453T} \textit{ten Brummelaar, T.A. et al.} First Results from the CHARA Array. II. A Description of the Instrument \textit{\apj} \textbf{628}, 453-465 (2005)

\bibitem{2006SPIE.6268E..55M} \textit{Monnier, J.D. et al.} Michigan Infrared Combiner (MIRC): commissioning results at the CHARA Array \textit{\spieconf} \textbf{6268}, 62681P-11 (2006)

\bibitem{2008ApJ...689L.137S} \textit{Stencel, R.E. et al.} Interferometric Studies of the Extreme Binary {$\epsilon$} Aurigae: Pre-Eclipse Observations \textit{\apjl} \textbf{689}, L137-L140 (2008)

\bibitem{2007Sci...317..342M} \textit{Monnier, J.D. et al.} Imaging the Surface of Altair \textit{\science} \textbf{317}, 342-345 (2007)

\bibitem{2005PASP..117.1255P} \textit{Pauls, T.A., Young, J.S., Cotton, W.D., Monnier, J.D.} A Data Exchange Standard for Optical (Visible/IR) Interferometry \textit{\pasp} \textbf{117}, 1255-1262 (2005)

\bibitem{2006SPIE.6268E..58I} \textit{Ireland, M.J., Monnier, J.D., Thureau, N.} Monte-Carlo imaging for optical interferometry \textit{\spieconf} \textbf{6268}, 62681T-8 (2006)

\bibitem{2008SPIE.7013E.121B} \textit{Baron, F., Young, J.S.} Image reconstruction at Cambridge University \textit{\spieconf} \textbf{7013}, 70133X-1 (2008)

\bibitem{2006SPIE.6268E..59L} \textit{Lawson, P.R. et al.} 2006 interferometry imaging beauty contest \textit{\spieconf} \textbf{6268}, 62681U-12 (2006)

\bibitem{2008ApJ...684L..95Z} \textit{Zhao, M. et al.} First Resolved Images of the Eclipsing and Interacting Binary {$\beta$} Lyrae \textit{\apjl} \textbf{684}, L95-L98 (2008)

\bibitem{1996ApJ...465..371L} \textit{Lissauer, J.J., Wolk, S.J., Griffith, C.A., Backman, D.E.} The epsilon Aurigae Secondary: A Hydrostatically Supported Disk \textit{\apj} \textbf{465}, 371-384 (1996)

\bibitem{2010:Hoard} \textit{Hoard, D.W., Howell, S.B., Stencel, R.E.} Taming the Invisible Monster: System Parameter Constraints for $\epsilon$ Aurigae from the Far Ultraviolet to the mid-Infrared \textit{\apj} in press (2010)

\bibitem{1997A&A...323L..49P} \textit{Perryman, M.A.C. et al.} The HIPPARCOS Catalogue \textit{\aap} \textbf{323}, L49-L52 (1997)

\bibitem{1970VA..W} \textit{K. O. Wright} The Zeta Aurigae stars \textit{Vistas in Astronomy} \textbf{12}, 147–182 (1970)

\bibitem{1978MNRAS.183..285B} \textit{Barnes, T.G., Evans, D.S., Moffett, T.J.} Stellar angular diameters and visual surface brightness. III - an improved definition of the relationship \textit{\mnras} \textbf{183}, 285-304 (1978)

\bibitem{2008ApJS..176..276V} \textit{van Belle, G.T. et al.} The Palomar Testbed Interferometer Calibrator Catalog \textit{\apjs} \textbf{176}, 276-292 (2008)

\bibitem{2002AA...386..492R} \textit{Richichi, A., Percheron, I.} CHARM: A Catalog of High Angular Resolution Measurements \textit{\aap} \textbf{386}, 492-503 (2002)





\end{thebibliography}
\end{document}